\documentclass[preprint,12pt]{elsarticle}

\usepackage{amssymb}

\usepackage{multirow}
\usepackage{graphicx}
\usepackage{graphics, slashed}
\usepackage{amssymb}
\def\ktresummation{$k_t$-resummation}
\def\ee{\end{equation}}
\def\be{\begin{equation}}
\def\eea{\end{eqnarray}}
\def\bea{\begin{eqnarray}}
\def\pp{$pp$}
\def\pbarp{${\bar p}p$}
\def\pipp{$\pi^+p$}
\def\pimp{$\pi^-p$}
\def\pip{$\pi p$}
\def\pipi{$\pi \pi$}
\def\pippim{$\pi^+\pi^-$}

\def\sigtot{\sigma_{tot}}

\def\sigtot{\sigma_{tot}}

\def\pimpim{\pi^-\pi^-}
\def\pippim{\pi^+\pi^-}
\def\x{cross-section}

\def\vecb{{\vec b}}

\def\nbar{{\bar n}}

\journal{Physics Letters B}

\begin{document}

\begin{frontmatter}



\title{Modeling pion and proton total cross-sections at LHC}


%%%%%%
\author[1]{A. Grau}
\ead{igrau@ugr.es}
\author[2]{G. Pancheri}
\ead{pancheri@lnf.infn.it}
\author[2]{O. Shekhovstova}
\ead{shekhovtsova@lnf.infn.it}
\author[3]{Y. N. Srivastava}
\ead{yogendra.srivastava@pg.infn.it}
\address[1]{Departamento de Fisica Teorica y del Cosmos, Universidad de Granada, Spain}
\address[2]{INFN Frascati  National Laboratories, P.O. Box 13, Frascati I00044,
    Italy}
\address[3]{Physics Department and INFN, University of Perugia, Perugia I06123, Italy}

\begin{abstract}
To settle the question whether the growth with energy is universal for different hadronic total cross-sections, we present results from theoretical models for $\pi p$, and  ($pp$,$p\bar{p}$) total cross-sections.
We show that present and planned experiments at LHC can differentiate between different models, all of which are consistent with presently available (lower energy) data . This study is also relevant for the analysis of those very high energy cosmic ray data which require reliable $\pi p$ total cross-sections as seeds. 
A preliminary study of the total $\pi\pi$ cross-sections is also made.     
\end{abstract}

\begin{keyword}
Hadronic  total cross-section \sep QCD  minijets \sep  Soft Gluon Resummation  \sep Froissart bound 
\end{keyword}

\end{frontmatter}

\section{Introduction}
\label{sec:introduction-not yet checked}
In the present work we describe theoretical predictions for  total pion-nucleon cross-sections  at LHC   using our  eikonal mini-jet model with soft gluon  \ktresummation \ \cite{Corsetti:1996wg,Grau:2009qx}.  We shall show  that at very high energy  there is considerable difference between current fits  and our QCD based model. We shall also present a preliminary estimate for the pion-pion cross-sections.

\begin{figure}[htbp]
\centering
\resizebox{\textwidth}{!}{%
\includegraphics{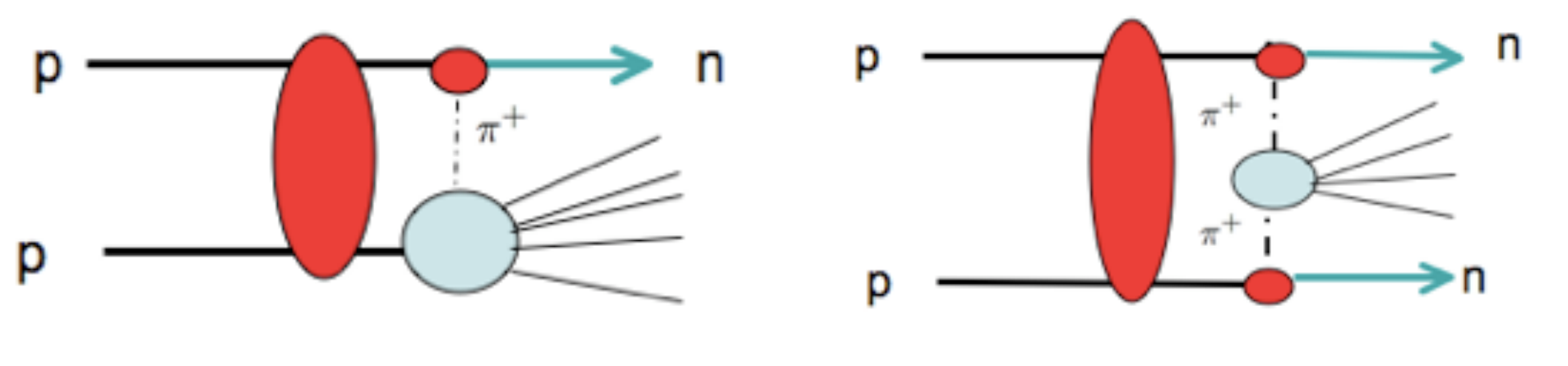}}
\caption{The pion exchange processes which would allow the measurement of $\pi^+p$ and $\pi^+\pi^+$ total cross-sections at LHC}
\label{piplhc}
\end{figure}

Recently a number of papers have appeared 
which point  both to the interest \cite{Petrov:2009wr} and the feasibility of  measuring  pion \x s at LHC \cite{Lebiedowicz:2010yb,Sobol:2010mu}.  The proposal for  LHC as a ``pion collider"  \cite{Petrov:2009wr}  and 
measurements of  $\pi N$ and even $\pi\pi$ scattering in the TeV range is based on the mechanism of pion exchange (single and double) and 
detection  of neutral particles  in the forward direction. Dedicated experiments such as the Zero Degree Calorimeter (ZDC) experiment\cite{Grachov:2008qg}, will place their detectors in the very forward region, a few micro radians from the beam and measure photons, $\pi^0$'s and neutrons. 
Detecting neutrons, total pion- proton and pion-pion cross-sections could be measured through the pion exchange mechanisms as shown in Fig.~\ref{piplhc}.

Extraction of the $\pi p$ total \x\ was suggested long time ago  \cite{Soding:1965nh}, rediscussed later in \cite{Ryskin:1997zz} and measured  in $\gamma p \rightarrow \pi^+\pi^- p$ at HERA \cite{Breitweg:1997ed}.
At CERN ISR, the measurement of the inclusive zero-angle neutron spectra gave strong  experimental support \cite{Flauger:1976ju} to the presence of an important charge exchange mechanism in the forward direction. This analysis was based on measurements of the inclusive differential cross-section 
for $pp\rightarrow nX$ at zero degree production angle and application of the triple Regge exchange mechanism to determine the exchanged trajectory $\alpha(t)$. In the kinematic limit of small $M^2/s$, but large $M^2,s$ one can write
\begin{equation}
\label{triple}
\pi E\frac{d^3\sigma}{d^3p}\approx \frac{d^2\sigma}{dt d(M_X^2/s)}=|G(t)|^2 (\frac{M^2}{s})^{1-2\alpha(t)}\sigma_{tot}(M^2,t)
\end{equation}
where $G(t)$ is the residue for the exchange of a reggeon between the proton and a neutron. One can interpret $\sigma_{tot}(M^2,t)$ as the total reggeon scattering cross-section at a CM energy $M$ and reggeon mass $|t|$.  Fits to the data indicated a value for the trajectory intercept at $t=0$ of $\alpha(0)=0.11\pm 0.15$, consistent with the pion trajectory. 
 Studies of the neutron spectra at HERA  \cite{Khoze:2006hw,Kaidalov:2006cw} revived interest  to the idea of using zero degree neutron detection  to study pion exchange processes. Recently,  forward neutron detection was examined  in \cite{Kopeliovich:2008da}, and  the extraction of the \x \ was discussed and a disagreement by a factor about 2 with the ISR data  claimed.  The data analyzed were forward neutron spectra from HERA \cite{Chekanov:2002pf,Chekanov:2007tv} and NA49 \cite{Varga:2004yu}.

 To have data for pion cross-sections  in the TeV range would be very interesting as it could help to effectively discriminate among models by studying more types of hadrons in the initial state in a similar energy range. 
Predictions for the total $pp$ \x \ at LHC energies vary from $~90\ mb$ to $140\ mb$ and more, with some models claiming even lower values and other somewhat higher ones.  LHC is expected to measure the total proton-proton cross-section with a precision of $5\%$ within the next 3 years, down to a precision of 1 \% when machine conditions at LHC will allow for it \cite{Anelli:2008zza}. Such measurements  should allow to discriminate among models.  An additional  strategy  relies on  comparing different initial states, such as photons or heavy ions, and their total \x \ value at very high energy. 
Valuable as they are however, data from photons or heavy ion collisions require additional modelling, and this may cloud the issue. Pions, on the other hand, should make a cleaner comparison to proton results.    Also, reliable total pion-nucleon cross-sections are needed for many cosmic ray analyses as well.

\section{The eikonal mini-jet model with soft gluon $k_t$-resummation}

We shall discuss total $\pi N$ cross-sections from a few GeV to about $100$ TeV CM energy range, and compare some current fits  to results from our eikonal mini-jet model implemented with soft gluon resummation in the infrared region \cite{Corsetti:1996wg,Grau:2009qx}. This model has two major advantages or differences with respect to other mini-jet models. First, it explicitly probes the large distance region through soft gluon resummation in the infrared. We  call it the Bloch-Nordsieck (BN) model because 
it relies on the role played by soft 
 quanta 
 resummation. We were inspired to build our model by the classic work of  Bloch and Nordsieck \cite{Bloch:1937pw} in electrodynamics, where they first  pointed  out that  only the emission of an infinite number of soft photons can lead to a finite result. In our model \cite{Grau:2009qx}, resummation and  implementation of soft gluon \ktresummation \ down into the gluon infrared momentum region constitute  the mechanisms through which the fast initial rise of all total cross-sections is trasformed into a smooth logarithimc  behaviour, which satisfies the Froissart bound. The second point in favour of our approach  is the calculation of the mini-jet \x s  using  actual Parton Density Functions (PDFs)  from available libraries, and inclusive of DGLAP evolution. Thus, we can make predictions for different processes simply by introducing the proper  PDFs in the formalism. 
Since the early observation of the rise of the total \x , parton-parton collisions were considered to be at the origin of the rise \cite{Cline:1973kv}.  QCD provides an obvious mechanism for this rise through the increasing number of low-energy perturbative gluon-gluon collisions, producing the so called {\it mini-jets} \cite{Pancheri:1984kc}. How to link these \x s to the total \x , without violating unitarity, was first done in the mid 80s  through the  eikonal mini-jet model \cite{Durand:1987yv,Durand:1988ax}.   Mini-jet models are used  in MonteCarlo simulation programs for very high energy collisions \cite{Sjostrand:1987su}, although there are some unsolved problems   concerning the elastic and diffractive components of the scattering, as   recently summarized in     \cite{Lipari:2009rm}. We shall not address these issues here  and  apply  the mini-jet model only to the total cross-section. For this purpose, we begin with the inelastic \x \ and then, using the fact that the real part of the scattering amplitude is 
expected to be
 small in the region of interest,  we construct the model for  the total \x 
 , with the eikonal expression
 
 \be
\sigma_{total}\approx 2\int d^2\vecb [1-e^{-\nbar (b,s)/2}]
\ee
 The simplest possible form for $\nbar (b,s)$ being  to write  $\nbar (b,s)=A(b)\sigma(s)$, one can see how   the    QCD mini-jet cross-sections  would  contribute to inelastic collisions at high energy as the driving term in the rise of $\sigtot$.    For scattering of particles A and B, an approximate expression  from  the low, $\sqrt{s}\approx   5\ GeV$,  to the highest energies is
  \be
  \nbar ^{AB} (b,s)=\nbar _{soft}^{AB}(b,s)+\nbar_ {QCD}^{AB}(b,s)=\nbar _{soft}^{AB}(b,s)
  +A_{BN}^{AB} (b,s) \sigma^{AB}_{\rm jet} (s,p_{tmin})
  \ee
where $p_{tmin}$ ($\sim 1-2 \ GeV$) is the  cut-off in the mini-jet \x \ $\sigma^{AB}_{\rm jet} (s,p_{tmin})$.    
 
As mentioned, and recently described in  \cite{Grau:2009qx}, we obtain   $\sigma^{AB}_{\rm jet} (s,p_{tmin})$, and hence the rise of the total \x  ,  through parton-parton scattering convoluted with 
 Parton Density Functions (PDFs) from PDFLIB, properly evolved with the scale of outgoing parton transverse momentum.  These are the mini-jet \x s in our model, and  different initial state particles can be studied simply by changing  the PDFs.
 
 To obtain a rate of increase with energy such as to describe both the early rise as well as the subsequent logarithmic behaviour, we introduce soft \ktresummation \ as the mechanism which generates an energy dependent acollinearity (and thus reduces the mini-jet \x  )  and write
 \be
 A_{BN}^{AB} (b,s)
 =\frac {
  e^{-h(b,q_{max})}
  }
{
\int d^2 \vecb   e^
{
-h(b,q_{max})
}
}=
\frac
{
exp
\{
-\frac{16}{3\pi}\int_0^{qmax} \frac{dk_t}{k_t}{\alpha_{eff}}(k_t)\ln 
(\frac{2q_{max}}{
k_t})
[1-J_0(bk_t)]\}
}
{\int d^2{\vecb}\   e^{-h(b,q_{max})}}
\ee
 where the function  $h(b,q_{max}) $
is obtained from soft gluon resummation techniques \cite{Corsetti:1996wg}, as indicated by the  the subscript {\it BN}  which recalls  the physics behind   this function. The function  $h(b,q_{max}) $ has a logarithmic  energy dependence through the scale $q_{max}$, which is proportional to $p_{tmin}$. 
 The excessive rise from  the minijets  is however reduced only by extending resummation to near zero soft gluon momenta, and this is accomplished through an {\it ad hoc} coupling, singular, but integrable, in place  of the asymptotic freedom expression for $\alpha_s$ in the soft gluon integral. With an effective coupling 
$\alpha_{eff}(k_t)\sim k_t^{-2p}$ for the single soft gluon transverse momentum  in the region $0 \le k_t \le \Lambda_{QCD}$
and $1/2<p<1$, we have shown in \cite{Grau:2009qx} how this expression for $A_{BN}^{AB} (b,s)$ introduces a strong cut-off in $b$-space and changes the violent rise of mini-jets into a smooth behavior in the total \x . Namely, we found that $\sigtot \sim (\ln s)^{1/p}$, a behaviour consistent with the limitations imposed by the Froissart bound.

To summarize, 
our model  for $\nbar _{QCD}^{AB}(b,s)$ is controlled by  three different momentum regions:
\begin{enumerate}
\item $p_t\ge p_{tmin}$,   for parton parton collisions, 
where a perturbative QCD description 
is applied,  with $p_{tmin}\sim (1-2)\ GeV$  kept fixed and independent of energy;
\item  $\Lambda_{QCD}\le k_t\le q_{max}$  for  
single soft gluons emitted from initial state quarks before the hard parton-parton collision,
 with \cite{Corsetti:1996wg} $ q_{max}\sim p_{tmin} \ln{\sqrt{s}/p_{tmin}}$; 
\item   $k_t\le \Lambda_{QCD}$ for ultrasoft gluons in  a region which is dominated by an effective  singular, but integrable, coupling of the gluons with the emitting quarks.
\end{enumerate}
The parameters which control the high energy rise of $\nbar _{QCD}^{AB}(b,s)$ are then 
\begin{itemize}
\item choice of PDFs (when different sets are available)
\item $p_{tmin}$
\item the singularity controlling parameter $p$.
\end{itemize}

Finally, 
  in this paper we use
  \be
  {\bar n}_{soft}^{AB}(b,s)=A_{FF}^{AB}(b)\sigma_{soft}^{AB}(s)
  \ee
  with the $b$-distribution given by the convolution of the form factors of the colliding particles.
 For protons and pions,
     this  gives 
 \bea
 A_{FF}^{pp}(b)&=&\frac{\nu^2 (\nu b)^3}{96\pi} {\cal K}_3(b \sqrt {\nu^2}), \ \ \   A_{FF}^{\pi \pi}(b)=\frac{q_0^3b}{4\pi}{\cal K}_1(bq_0) \nonumber \\
 A_{FF}^{\pi p}(b)&=& \frac{1}{4\pi} \frac{\nu^2 q_0^2}{q_0^2-\nu^2} \Bigl[\nu b {\cal K}_1(\nu b)- \frac{2\nu^2}{q_0^2-\nu^2}[{\cal K}_0(\nu b)-{\cal K}_0(q_0 b)]\Bigr] 
 \label{eq:aff}
 \eea
 with $\nu^2=0.71 \ GeV^2$ for the proton and $q_0=0.735\ GeV$ for the pion.
 At   low energy, we parametrize the \x \ in the eikonal  as 
\be
  \sigma_{soft}^{AB}(s)=A_0+\frac{A_1}{ E_{lab}^{\alpha_1}}-\frac{A_2}{ E_{lab}^{\alpha_2}}
  \label{eq:sigmasoft}
\ee  
and $E_{lab}=\frac{s-m_A^2-m_B^2}{2m_A}$ for particle B on fixed target A. In Eq.~(\ref{eq:sigmasoft}),
the coefficients $A_i$ and the exponents $\alpha_i$ are in principle different for each different process. Following our analyses in \cite{Grau:1999em,Godbole:2004kx}, we  chose the set of high energy parameters given by GRV densities and set \{1,5,4\} \cite{Gluck:1991ng}, $p_{tmin}=1.15 \ GeV$ and $p=0.75$. We then applied  Eqs.~\ref{eq:aff},\ref{eq:sigmasoft} to determine the low energy parameters for   \pp \ and \pbarp \ scattering and   performed  a best fit  to the overall data set. The results of the fit  for the low energy parameters 
are shown in Table ~\ref{Table:protons}. 
\begin{table*}
\caption{Results of the fit to $pp$ , $p\bar p$  data}
\label{Table:protons}
\begin{center}
  \begin{tabular}
{|c|p{7cm}|p{7cm}|}
\hline
{Process}&\multicolumn{2}{l|}
{Fit}\\
\hline
\multirow{4}{*}{$pp$}&\multicolumn{2}{l|} {$A_0 = (48.20 \pm 0.19)$ mb }
      \\ &\multicolumn{2}{l|} { $A_1 = 101.66 \pm 16.35$  \hspace{0.2cm}
$\alpha_1 = 0.99 \pm 0.13$}  
      \\ &\multicolumn{2}{l|} { $A_2 = 27.89 \pm 4.78$ \hspace{0.2cm} $\alpha_2 = 0.59 \pm 0.04$} \\
    & \multicolumn{2}{l|} { $\chi^2 = 154.1/(102+5-1)$} \\
\hline
\multirow{4}{*}{$p\bar p$}&\multicolumn{2}{l|} { $A_0 = (47.86 \pm 2.47)$ mb} 
      \\ & \multicolumn{2}{l|} {$A_1 = 132.07 \pm 32.89 $  \hspace{0.5cm}
$\alpha_1 = 0.69 \pm 0.14$  }
      \\&\multicolumn{2}{l|} {  $A_2 = 0.82 \pm 0.31 $  \hspace{0.5cm} $\alpha_2 = 0.52 \pm 0.07$} \\
     &\multicolumn{2}{l|}{$\chi^2=24.65/(31+5-1)$} \\
\hline \hline
 \end{tabular}
\end{center}
\end{table*}
%%%%%

\begin{figure}
\centering
\vspace{-4cm}
\hspace{-5mm}
\resizebox{0.8\textwidth}{!}{
\includegraphics{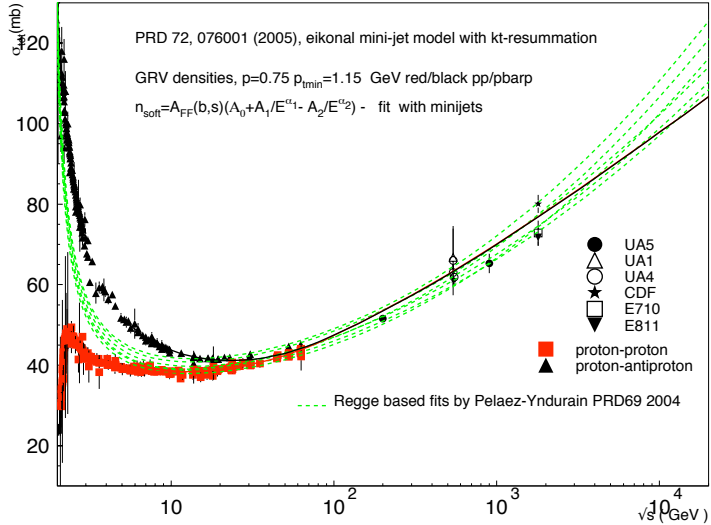}}
\vspace{-4cm}
\caption{Results of our model for \pp \ and \pbarp \ scattering compared with data and with Regge-based fits by Pelaez and Yndurain \cite{Pelaez:2003ky}.  Fits for  low energy parameters are in the variable $E\equiv E_{lab}$.}
\label{fig:august2protons_olgafit_py}
\end{figure}

Our curves are shown 
in Fig.~\ref{fig:august2protons_olgafit_py}  together with fits by Pelaez and Yndurain  [PY] \cite{Pelaez:2003ky} for the combination $(pp+{\bar p}p)/2$. These fits are part of a global fit to $\pi \pi, \pi N, KN $ and $pp, {\bar  p}p$. The PY fits  
 follow Regge theory constraints in the low energy region and incorporate    
a high energy term which follows a more stringent limit  \cite{Yndurain:2003vk} than the one imposed by the Froissart bound,  i.e. 
\be
\sigma_{PY}=
a_0+a_1s^{\beta_1}+a_2s^{\beta_2}+B_{PY} \ln^2{\frac{s}{s_1\ln^{7/2}{s/s_2}}}
\ee
 In our model, on the other hand, and as discussed before, we do not impose {\it a propri} the presence of any  logarithmic term, rather we can see  that at high energy the model naturally develops a logarithmic behaviour, not present in the low energy terms.

Having thus recalled the basic formulae which we shall use   for extension of our model to pion processes,  we describe in the next section what the model predicts for pion-proton cross-sections in the very high energy region, accessible through LHC as well as cosmic ray experiments,  and   compare our predictions with available parametrizations.

\section{Phenomenology of $\pi p $ total \x s : comparing models and their predictions into the  TeV range }
In this section we apply our model, as described in the previous section, to  $\pi p$ scattering, comparing it with existing data, models and fits.  Data  from  fixed target experiments \cite{Amsler:2008zzb} are available up to $\sqrt{s}=36.7\ GeV$ for   \pimp \ and  $\sqrt{s}=25.3\ GeV$ for \pipp .  A second set of data in the range $9.4\ GeV \le \sqrt{s} \le 70\ GeV$  is the one  obtainable   from the charge exchange mechanism \cite{Petrov:2009wr} described in  the introduction.
We examine   
four  predictions for $\pi^+ p$ total cross-section at LHC energies, namely: 

\begin{itemize}
\item a Regge-Pomeron fit from Donnachie and Landshoff \cite{Donnachie:1992ny}
\begin{equation}
\sigma_{\pi^+p}(mb)=13.63 s^{0.0808}+27.56s^{-0.4525}
\end{equation}
 noting  that for  $\pi^-p$ 
 the coefficient of the second (``Regge" ) term is changed from 27.56 to 36.02; 
\item the fit from the COMPETE/PDG 2008 collaboration \cite{Amsler:2008zzb} given as 
\begin{equation}
\sigma_{\pi^+p}=Z^{\pi p}+B\ln^2 (\frac{s}{s_0})+Y_1^{\pi p}(\frac{s_1}{s})^{\eta_1}-Y_2^{\pi p}(\frac{s_1}{s})^{\eta_2} 
\label{eq:fitcudellpip}
\end{equation}
with
$Z^{\pi p} = 20.86\ {mb}, \ B = 0.308\  {mb}, \ Y^{\pi p}_1 = 19.24\ {mb}, \ Y^{\pi p}_2 = 6.03\ {mb}, \ \eta_1 =0.458,\ \eta_2 = 0.545, \ s_1 = 1\ {GeV}^2, \ \sqrt{s_0}= 5.38\ {GeV}$
\item a fit by  Block and Halzen \cite{Block:2005pt} with a  functional expression similar to  the one from PDG but with an extra term linear in $\ln{s}$, 
\begin{equation}
\sigma^{ab}=c_0+c_1\ln{(\nu/m_\pi)}+c_2\ln^2{(\nu/m_\pi)}+
\beta (\nu/m_\pi)^{\eta_1}+\delta (\nu/m_\pi)^{\eta_2}
\end{equation}
with numerical coefficients for  $\pi^+ p$ given by 
$
c_0=20.11\  {mb}, \ c_1 = -0.921  \ {mb},\  c_2 = 0.1767  \ {mb}, \  \beta=54.4\  {mb}, \ \delta = -4.51\  {mb}, \ 
\eta_1 = -0.5, \ \eta_2 = -0.34 $  and $\nu$ the laboratory energy;
\item the eikonal mini-jet model with initial state soft gluon $k_t$-resummation described in the previous section, with GRV density functions for both the pion \cite{Gluck:1991ey} and the proton \cite{Gluck:1991ng}. 
\end{itemize}

We shall now enter more into the details of our calculation. We 
 proceed to study \pipp \ and \pimp  \ by fitting  the low energy part of the eikonal function, ${\bar n}_{soft}^{AB}$, 
with $A_0^{\pi p}$ either fixed to $2/3 \ A_0^{pp}$, following the Additive Quark Parton Model (AQPM) rule,   or free to vary, and  the high energy part ${\bar n}_{QCD}^{AB}$ computed  with   the  same  parameters used  for \pp \ or \pbarp  , except that the PDFs are now those for pions. Thus, we used  $p_{tmin}=1.15\ GeV, p=0.75$, and GRV densities for pions and protons. Then, through a fit which includes the minijet contribution in the eikonal, we  determine  the low energy parameters entering ${\bar n}_{soft}^{\pi^\pm p}$. The low energy part of the \x \ is as before obtained through 
 Eqs.~\ref{eq:aff} and \ref{eq:sigmasoft}. Notice that this procedure includes the rise due to mini-jets even when their contribution is too small to be actually detected.

The scope of this exercise is twofold: to see whether the AQPM rule works for the constant term of the eikonal function, and obtain predictions for the high energy behaviour based on the same parameter set as in \pp \ or \pbarp . We do not actually know whether the high energy parameters should be the same for all processes or be energy independent, but  keeping them fixed, as we do,  can give   information about their universality.
The result for these two different ways to study $\pi p$ scattering at high energy is then examined by the $\chi^2$ for the two different cases,  \pimp \ and  \pipp . We show the results of the fit in Table ~\ref{tab:pionproton}.

%%%%%
\begin{table*}
\caption{Results of the fit to 
 $\pi^+ p$, $\pi^- p$
data.
 } 
\label{tab:pionproton}
\begin{center}
  \begin{tabular}
{|c|p{7cm}|p{7cm}|}
\hline
 \multirow{2}{*}{Process} &
Fit 1, fixed value of $A_0$ & Fit 2 
      \\ & equal to $A_0= \frac{2}{3}A_0(pp)$ & free value of $A_0$\\ \hline
 \multirow{4}{*}{$\pi^+ p$}  & $A_0 =32$ mb & $A_0 = (28.5 \pm 0.13)$ mb 
      \\ & $A_1 = 37.9 \pm 2.4$  \hspace{0.5cm} $\alpha_1 = 0.46 \pm 0.02$ & 
         $A_1 = 80.8 \pm 1.1$ \hspace{0.5cm} $\alpha_1 = 0.52 \pm 0.05$ 
      \\  & $A_2 =24.6 \pm 0.96 $  \hspace{0.5cm} $\alpha_2 = 0.20 \pm 0.01$ & 
           $A_2 = 58.9 \pm 0.95$ \hspace{0.5cm} $\alpha_2 = 0.40 \pm 0.004$
       \\ & $\chi^2 = 70/(50+4-1)$ &  $\chi^2 = 69.5/(50+5-1)$      
\\ \hline
 \multirow{4}{*}{$\pi^- p$} & $A_0 =32$ mb & $A_0 = (27.4 \pm 0.1)$ mb 
      \\ & $A_1 = 37.4 \pm 4.5$  \hspace{0.5cm} $\alpha_1 = 0.41 \pm 0.03$ & 
        $A_1 = 84.4 \pm 0.7$ \hspace{0.5cm} $\alpha_1 = 0.57 \pm 0.02$ 
      \\  & $A_2 = 20.8 \pm 4.8 $ \hspace{0.5cm}  $\alpha_2 = 0.16 \pm 0.02$ & 
           $A_2 = 55.5 \pm 1.7$ \hspace{0.5cm} $\alpha_2 = 0.50 \pm 0.02$
       \\ & $\chi^2 = 193/(95 + 4 - 1)$ & $\chi^2 = 197/(95+5-1)$  \\
    \hline
     \end{tabular}
\end{center}
\end{table*}
 We find that one obtains a good fit to the data in either of the two cases, 
  $A_0$ free or given by the AQPM rule of 2/3.
Before proceeding further, we  notice a   difference between our result for the low energy fit and other fits: our constant term $A_0$ can fit the data with a value  $\approx 2/3 $ of the constant term used  in the eikonal for \pp \ and \pbarp  ,
while the other fits (PDG and BH), have a ratio of the constant terms $< 2/3$. This may be  due to the fact that they assume the contribution  of a $\ln^2 {s}$ or $\ln {s}$ term throughout the entire  energy region , whereas in our model the logarithmic behaviour  arises naturally 
through the minijets and the \ktresummation \ effect mentioned in the previous section and described in \cite{Grau:2009qx}. Such terms, although present in the fittimg procedure, start contributing  only for   $\sqrt{s}\ge 10 \ GeV$.

\begin{figure}
\begin{center}
\par
\parbox{1.0\textwidth}{\hspace{-0.5cm}
\vspace{-.5cm}
\includegraphics[width=0.6\textwidth,height=0.5\textwidth]{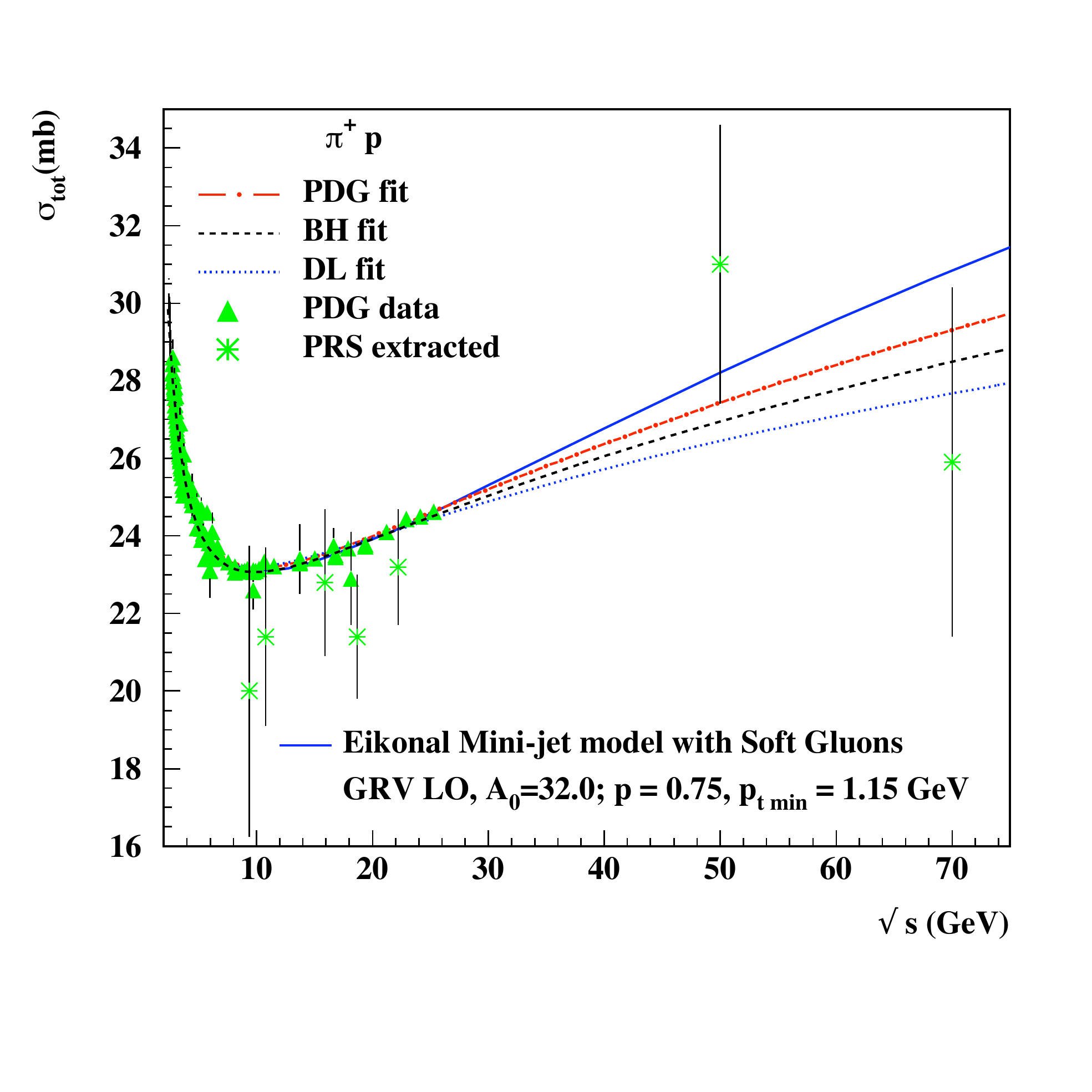}
\includegraphics[width=0.6\textwidth,height=0.5\textwidth]
{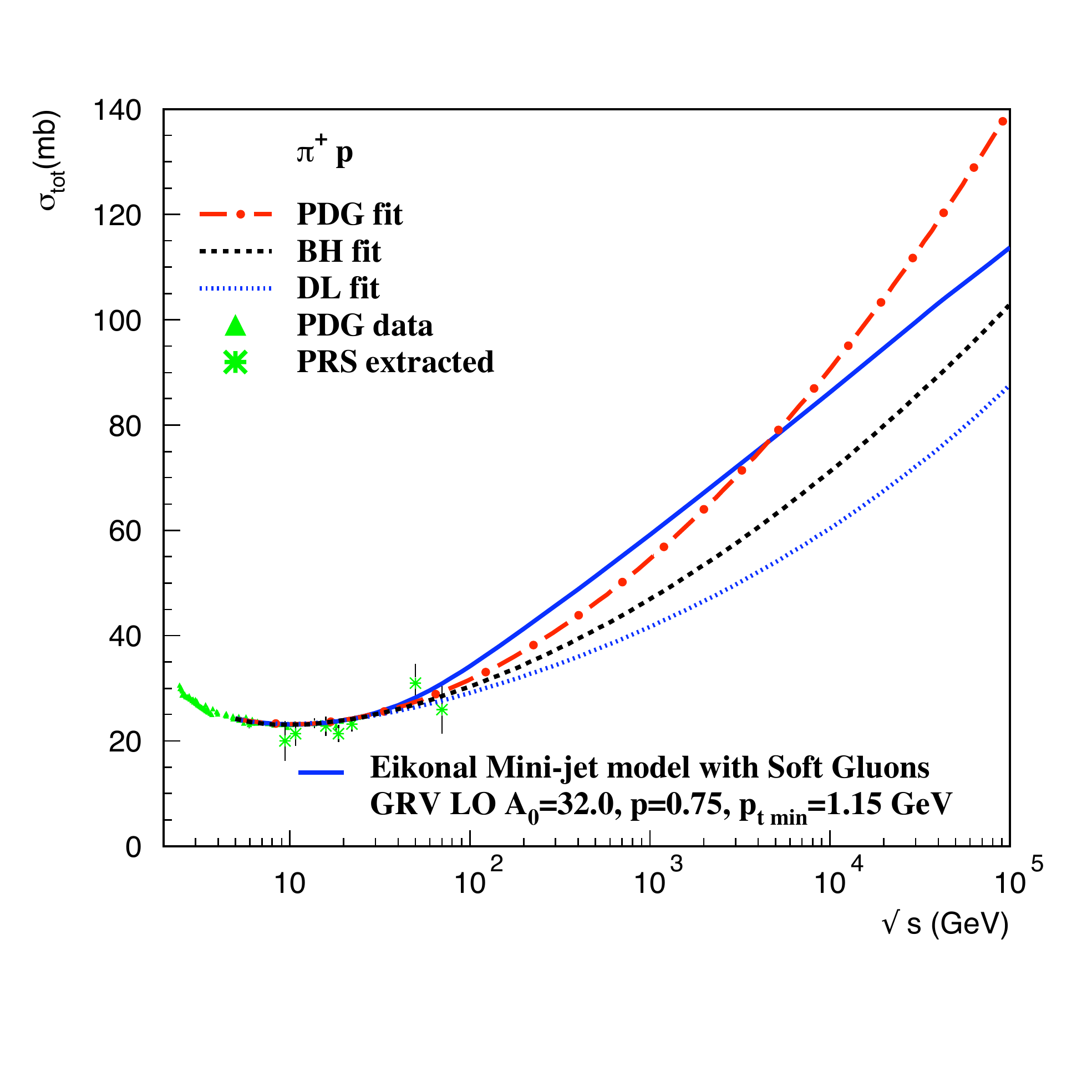}}
\caption{In the left panel we show the fit  for $\pi^+ p$ channel in the low energy region when   the parameter $A_0$  is fixed to the value $2/3\ A_0^{pp}$, and with    values of the low energy  parameters as in Table \ref{tab:pionproton}.  At right, we show the full energy range of interest.  
}\label{fit_pi+p}
\end{center}
\end{figure}

In Fig.~\ref{fit_pi+p}  we  compare   the  fits  labelled DL, BH and PDG   with existing data for \pipp \ and  with the results from our model.
 We show  a comparison in the low energy region, where there are data, and the high energy predictions.   The points labelled PRS are from \cite{Petrov:2009wr} and have been extracted from actual data, BH is the fit from \cite{Block:2005pt}, and DL from \cite{Donnachie:1992ny}. 
We show the case $A_0$ fixed according to the AQPM model, but, as mentioned,  there is no discernible difference in the results when   $A_0$ is free to vary. 

 We notice that there is a difference of almost a factor 2 among the asymptotic limit of different curves, indicating the interest of this exercise. This difference is of course due to the fact that data for $\pi p$  extend only up to the beginning of the rise and fits cannot really adequately determine the asymptotic behaviour. The situation here is quite different from  the  \pp \ case  where data constrain the LHC value to within a 10-20\% range, an uncertainty 
carried on from the Tevatron data.
The 
 extension of  our investigation to energies in the very high cosmic ray region, namely up to $\sqrt{s}\sim 100\ TeV$ shows that at this energy there is no difference, as expected, between $\pi^+p$ and $\pi^-p$ \x s, but an even larger difference among different fits and our model is observed. We  reproduce  these results 
 in Table ~\ref{tab:100}.
\begin{table}
\caption{ Total $\pi^{\pm}p$ \x \ at $\sqrt{s}=100\ TeV$ }
\label{tab:100}
\centering
\begin{tabular}{|c||c|c|}
\hline\noalign{\smallskip}
Model&Value in mb&Value in mb\\ 
            & $\pi^-p$      &$\pi^+p$\\ \hline
PDG&139.9&139.9\\
BH&102.89& 102.88\\
DL&87.59& 87.59\\
EMM (fixed A0)&113.72&113.75\\
EMM (free A0)&113.28&113.40\\
\noalign{\smallskip}\hline
\end{tabular}
\end{table}

\section{$\pi \pi$ scattering}
The possibility to measure elastic $\pi^+\pi^+$ scattering at LHC has recently been discussed at some length \cite{Sobol:2010mu}. A related exclusive cross-section measurement for the process $pp\rightarrow\ nn \pi^+\pi^+$ is discussed in \cite{Lebiedowicz:2010yb}. At a $p\bar{p}$ collider (such as the Tevatron), one has in principle the possibility of obtaining the cross-section for the channel $\pi^+\pi^-$ through the process $p\bar{p}\rightarrow\ n\bar{n} \pi^+\pi^-$. One may even entertain the possibility of measuring -at the LHC- the neutral $\pi^+\pi^-$ channel via the measurement $pp \rightarrow\ n  \Delta^{++} \pi^+ \pi^-$ and the $\pi^+\pi^o$ channel via the reaction $pp \rightarrow\ n  \Delta^{+} \pi^+ \pi^o$.

A good  overview of the available total    cross-section data for like-sign and opposite sign $\pi\pi$ at low energy is  summarized in \cite{Pelaez:2003ky}.
There are no data for the  process $\pi^+\pi^+$ but isospin invariance tells us that the \x \ is the same as the one for $\pimpim$, for which there exist data which   we shall then use for a fit. However,  unlike the other scattering processes we considered until now, data for the $\pi \pi$ channels \cite{PhysRevLett.18.273,PhysRevD.7.661,Robertson:1973tk,Hanlon:1976ct,Hoogland:1977kt,Abramowicz:1979ca} do not extend into the region where minijets start playing a role and, in addition, these data are in some contradiction with each other and have large errors. This makes fits and the error analysis particularly difficult,   and the extension to higher energies uncertain.  We shall return to a detailed investigation of this point in subsequent work.

 With the above {\it caveats}, we follow the strategy presented before and  perform a fit to the data to determine the low energy parameters. Starting with $\pi^+\pi^-$  and leaving $A_0$ free, we obtain
\bea
A_0^{\pi^+\pi^-}&=& (22.0 \pm 9.8)\ mb,  \ A_1= 100.0 \pm 89.8 ,\alpha_1=0.74 \pm 0.66, \\
 A_2 &=& 27.8 \pm 49.1,\alpha_2=0.27 \pm 0.33 
\eea
with $\chi^2=6.93/(17+5-1)$. 
The $\pi^+ \pi^-$ fit done  leaving $A_0$ free  gives a central value  $A_0^{\pi \pi}=4/9\ A_0^{pp}$ as expected from the AQPM rule. However,    the parameters for $\pi\pi$ are determined with large errors.
In this paper, we have presented results from our
model using the best mean values for the fitted parameters. The 
encouraging results which we obtain justify {\it a fortiori} such a
procedure.  This result for $A_0^{\pippim}$ being close  to what one would obtain through the AQPM rule, we then proceed to perform the fit for $\pimpim$.
    The results of the fit to the low energy parameters for $\pi^-\pi^-$ 
  are given in  Table~\ref{tab:pimpim}.
\begin{table*}
\caption{Results of the fit to 
 $\pi^-\pi^-$  data.}
\label{tab:pimpim}
\begin{center}
  \begin{tabular}
{|c|p{7cm}|p{7cm}|}
\hline
\multirow{2}{*}{Process} &
Fit 1, fixed value of $A_0$ equal to  & Fit 2 
      \\ &  free fit value for $A_0(\pi^+\pi^-)$ & free value of $A_0$\\ \hline
 \multirow{4}{*}{$\pi^- \pi^-$} & $A_0 =22.0$ mb & $A_0 = (12.3 \pm 0.2)$ mb 
      \\ & $A_1 = 85.5 \pm 1.1$ \hspace{0.5cm}  $\alpha_1 = 0.08 \pm 0.08$ & 
        $A_1 = 80.2 \pm 2.1$ \hspace{0.5cm} $\alpha_1 = 0.53 \pm 0.004$ 
      \\  & $A_2 = 99.7 \pm 0.9 $ \hspace{0.5cm}  $\alpha_2 = 0.08 \pm 0.07$ & 
           $A_2 = 99.9 \pm 0.02 $ \hspace{0.5cm} $\alpha_2 = 0.57 \pm 0.001$
       \\ & $\chi^2 = 53.19/(14+4-1)$  & $\chi^2 = 49.19/(14 + 5 - 1)$  \\
    \hline
     \end{tabular}
\end{center}
\end{table*}
   Notice that, both for $\pimpim$ as well as for $\pippim$,   making the fit in the variable $s=M^2_{\pi\pi}$ leads  to very similar results as those for fits in the variable $E_{lab}$.
The small error from data reported by \cite{Abramowicz:1979ca} compared to those in \cite{Robertson:1973tk}  makes the fit leaning  towards the lower numerical results. 
 
 As in the case of $\pi p$ scattering,   the high energy part, mini-jets and soft gluon resummation, is then  calculated with the same set of parameters as for \pp \ and \pbarp \ and $\pi p$, namely GRVLO densities, $p_{tmin}=1.15\ GeV$ and $p=0.75$. For $\pimpim$, at high energy, the effect of the two different fits of the low energy data results in $\sim 1\  mb$  difference in the total \x \ at high energy.

\begin{figure}
\centering
\vspace{-20mm}
\hspace{-5mm}
\begin{minipage}{0.5\linewidth}
\hspace{-10mm}
\includegraphics[width=1.2\textwidth]{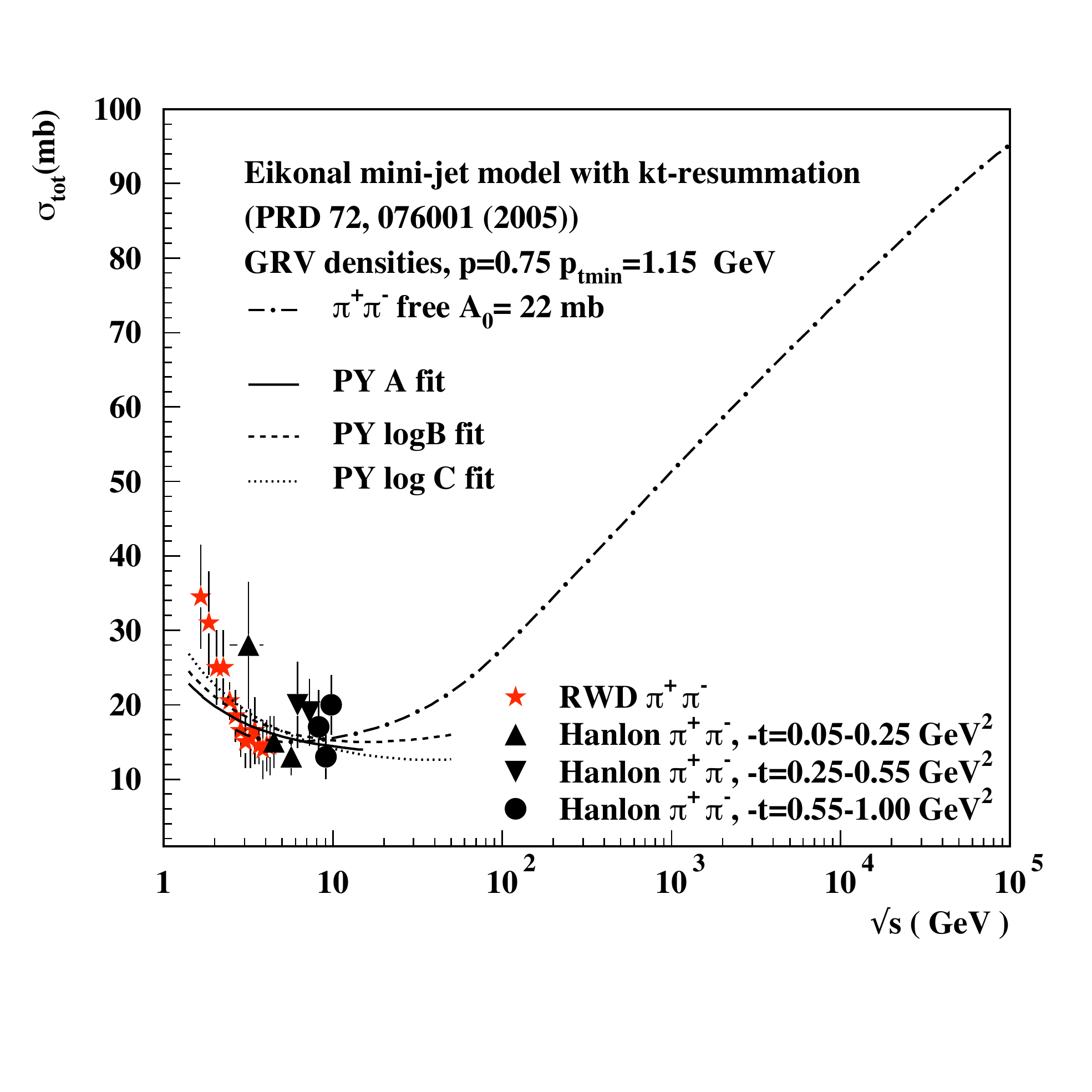}
\end{minipage}
\begin{minipage}{0.5\linewidth}
\includegraphics[width=1.2\textwidth]{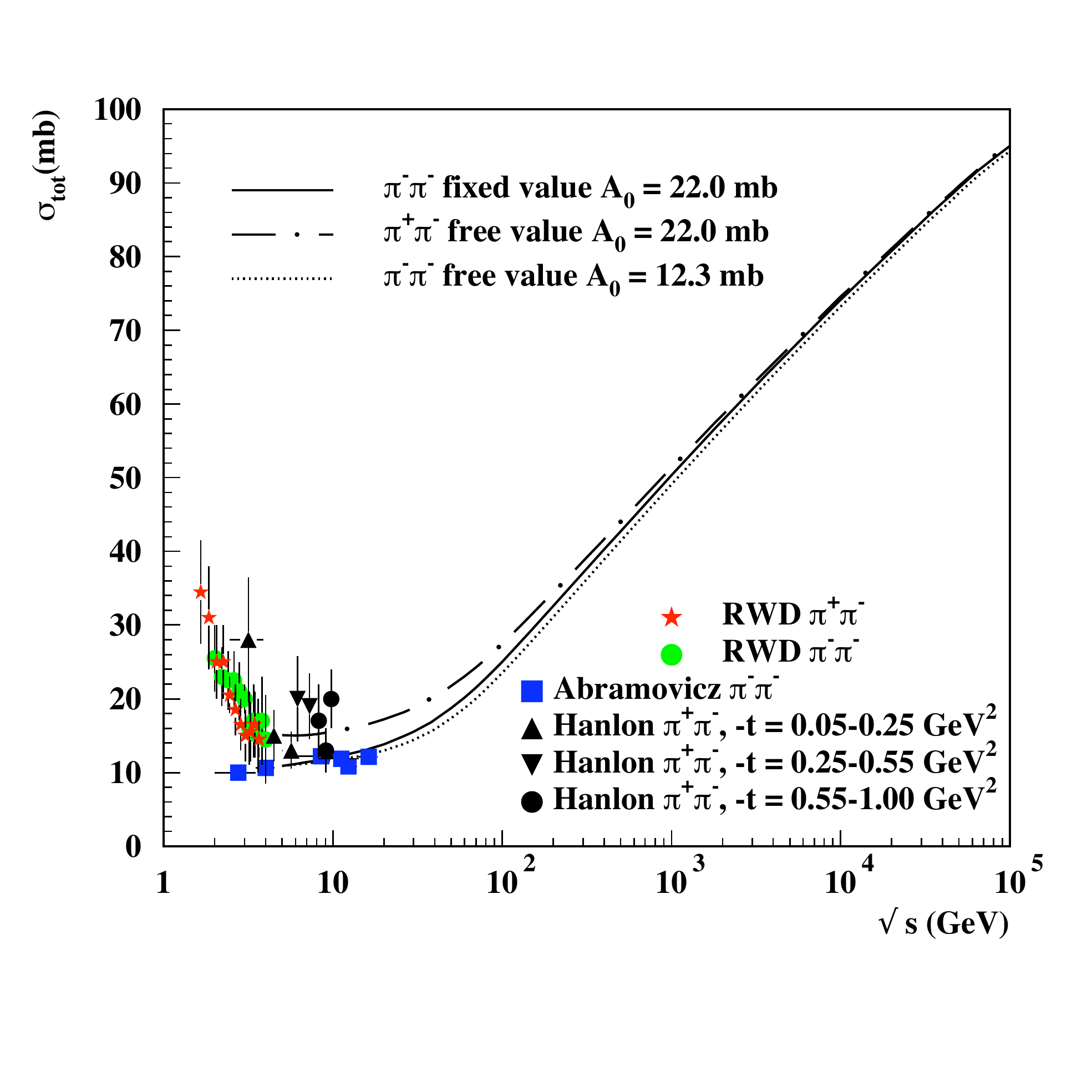}
\end{minipage}
\label{fig:pipi}
\vspace{-1cm}
\caption{ Predictions  for $\pi^+\pi^-$ and $\pimpim$ scattering using our eikonal minijet model are shown. 
 At left our results are compared with the central values of 
 PY \cite{Pelaez:2003ky}. For $\pi^+\pi^-$ data , the   constant term in the fit is given by $A_0^{\pi \pi}=22 \ mb \approx 4/9 A_0^{pp}$. Our predictions for $\pimpim$ using both free and fixed $A_0$ are also shown. RWD data  are from \cite{Robertson:1973tk},   Abramowicz et al. data are from  \cite{Abramowicz:1979ca},  Hanlon et al. data are from \cite{Hanlon:1976ct}.}
\end{figure}

Our results for $\pi^+\pi^-$ and $\pimpim$ 
 are shown in  Fig.~\ref{fig:pipi}, where
   we reproduce the data we have  found in the literature. In the left panel 
  we compare data for $\pi^+\pi^-$ with our model predictions as well as with the Regge based parametrizations by Pelaez and Yndurain \cite{Pelaez:2003ky}. We notice the difference in the energy behaviour between our curves and the PY fits. At low energy the difference may be due to our fitting data from $\sqrt{s}=2.5 \  GeV$ only.  This follows from the fact that we always include in the fits the mini-jet contribution, and the relative PDFs cannot be used at lower energy values. At  the same time   lack of data in the region where the rise is expected to start reduces the predictive power of fits. This difference, between low energy fits and high energy behaviour from models, makes a measurement of this process particularly interesting.

\begin{figure}
\centering
\vspace{-5cm}
\resizebox{0.8\textwidth}{!}
{\includegraphics{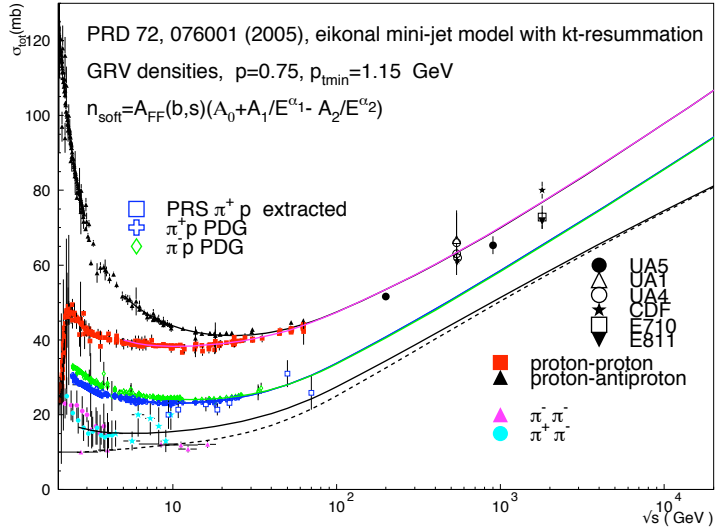}}
\vspace{-4cm}
\caption{Our curves for \pp \  , \pbarp \ , $ \pi^{\pm} p$ and $\pi \pi$ using the same high energy parameters and densities, and different low energy parametrizations, as described in the text. For $\pimpim$ the (dashed) curve corresponds to fixed $A_0$=22\ mb. }
\label{fig:august12pionsprotons}
\end{figure}

Finally  we  compare all the six processes we have examined in this paper. We  give the numerical results for some representative energy values 
in Table~\ref{tab:tableall}
 and then show the results in an overall figure.
\begin{table}
\caption{Numerical values obtained from our model for proton and pion total cross-sections in mb at selected  cm energy values,  using  central  values for the low energy parameters.}
\label{tab:tableall}
\vspace{5mm}
\begin{center}
\begin{tabular}{|c|c|c|c|c|c|c|c|c|}
\hline
$\sqrt{s}$&$pp$  &$p{\bar p}$&$\pi^+ p$            &$\pi^-p$         &$\pi^+ \pi^-$       &$\pi^-\pi^-$&$\pi^-\pi^-$ &${\cal R}$\\
(GeV)      &       &           &free $A_0$ &free $A_0$& free $A_0$ &fixed $A_0$& free $A_0$& \\ \hline
5		 & 40  & 51.9    & 24.3           & 26            &15               &10.9           &10.8 &1.06 \\	
100         &47.4 & 47.5    &33.8           &34             &27.5            & 25             & 23.1 & 0.94\\
1000       &70.1 & 70    &58.5          &59.0           &51.3            &   50.3        &48.9 &0.98\\
10000     & 97.9&97.8  & 85.7         & 86.1         & 74.6            & 74.1          &73.1 &1.02 \\
\hline
\end{tabular}
\end{center}
\end{table}
In Table ~\ref{tab:tableall},  the ratio ${\cal R}$  is calculated  from
\be 
{\cal R}=\frac{
<\sigma_{tot}^{\pi\pi}>_N}{<\sigma_{tot}^{\pi\pi}>_{model}}
\ee 
and gives  the results of a "factorization" exercise, namely whether our model satisfies  factorization
for \pipi \ scattering, with  $<\sigma_{tot}^{\pi\pi}>_{model}$ calculated through the values in Table~\ref{tab:tableall}
as the average of $\pi^+\pi^-$ and $\pi^-\pi^-$. For the latter, we take the average between free and fixed $A_0$ parametrizations.
 Factorization is tested by comparing the model results with the factorized expression obtained from   
\be
<\sigma_{tot}^{\pi\pi}>_N=\frac{
(\sigma_{tot}^{\pi N})^2
}{
\sigma_{tot}^{NN}
}
\ee
with 
\be
\sigma_{tot}^{\pi N}=\frac{\sigma^{\pi^+p}+\sigma^{\pi^-p}}{2}, \ \ \ \ \  \sigma_{tot}^{NN}=\frac{\sigma^{pp} +\sigma^{p{\bar p}}}{2}
\ee
 In  Fig,~\ref{fig:august12pionsprotons} we collect the results from our model combining together all proton and pion total \x s. 

\section{Conclusion}
Our minijet model for total cross-sections for (i) $NN$ (and $N\bar{N}$), (ii) $\pi N$ and (iii) $\pi \pi$ exhibits some  general characteristics. In the low-to-medium energy range the quark counting rule works quite well (that is, the ratio $1:2/3:4/9$ for the three cases holds). At higher energies, the three cross-sections appear to rise roughly at a similar rate. This analysis confirms the general belief that the quark model is valid at lower energies,  whereas, in the asymptotic domain, it gives strong support that our model based on minijets with soft gluon resummation provides an adequate description of all hadronic total cross-sections. Experimentally, the data are most precise for the $NN$ case and less so for the $\pi N$ case. Data for pions either do not extend to the region where the rise is well established, as in the case of \pip \ , or   are insufficient to make a good fit at low energy, as is the case for \pipi \ . This does not allow for reliable fits at very high energy. On the other hand, models can be tested. Our model for the high energy part has the virtue of being of straightforward application, when substituting pions for protons. Once the values for the parameters for $p$ and $p_{tmin}$ are chosen,  the PDFs  available for the particles under consideration provide  the mini-jet \x \ and the soft scale $q_{max}$  for the $b$-distribution. Clearly, measurements at LHC for pion reactions would be most useful.
\section*{Acknowledgments}
O.S. is  grateful to S. Eydelman of Novosibirsk for enlightnening conversations about  pion data, and G.P.  thanks  M. Murray, CMS,  U. of Kansas, for discussions about the planned ZDC measurements and R. Pelaez and Jacobo Ruiz de Elvira from U. Madrid for helping us with data from the Pelaez and Yndurain fits..
G.P. gratefully acknowledges  the hospitality of the Center for Theoretical Physics of MIT and Brown U. Physics Department. Y.S. would like to thank the Physics Department for his stay as an Emeritus Professor at Northeastern University, Boston. Work partially supported by the Spanish MEC 
(FPA2006-05294 and FPA2008-04158-E/INFN) and by Junta de Andaluc\'\i a 
(FQM 101).
%%%%%%%%%%
% \bibliographystyle{model1a-num-names}
 \bibliographystyle{model1-num-names}
  \bibliography{pion-short-august24}
 \end{document}